\documentclass[journal]{IEEEtran}
\usepackage{graphicx}
\usepackage{amsmath, amssymb}
\usepackage{caption, subcaption}
\usepackage{float}
\usepackage{multirow}
\usepackage{hyperref}
\begin{document}
\title{A Fast 3D CNN for Hyperspectral Image Classification}
\author{Muhammad Ahmad,
\thanks{M. Ahmad is with the Department of Computer Engineering, Khwaja Fareed University of Engineering and Information Technology, Rahim Yar Khan, 64200, Pakistan, E-mail: mahmad00@gmail.com}
}
\markboth{Preprint Submitted to arXiv, April~2020}
{M.Ahamd, A Fast 3D CNN for HSIC}
\maketitle
\begin{abstract}Hyperspectral imaging (HSI) has been extensively utilized for a number of real-world applications. HSI classification (HSIC) is a challenging task due to high inter-class similarity, high intra-class variability, overlapping, and nested regions. A 2D Convolutional Neural Network (CNN) is a viable approach whereby HSIC highly depends on both Spectral-Spatial information, therefore, 3D CNN can be an alternative but highly computational complex due to the volume and spectral dimensions. Furthermore, these models do not extract quality feature maps and may underperform over the regions having similar textures. Therefore, this work proposed a 3D CNN model that utilizes both spatial-spectral feature maps to attain good performance. In order to achieve the said performance, the HSI cube is first divided into small overlapping 3D patches. Later these patches are processed to generate 3D feature maps using a 3D kernel function over multiple contiguous bands which persevere the spectral information as well. Benchmark HSI datasets (Pavia University, Salinas and Indian Pines) are considered to validate the performance of our proposed method. The results are further compared with several state-of-the-art methods.\end{abstract}
\begin{IEEEkeywords}
3D Convolutional Neural Network (CNN); Kernel Function; Classification; Hyperspectral Images (HSI);
\end{IEEEkeywords}
\IEEEpeerreviewmaketitle
\section{Introduction}
\label{Sec.1}

\IEEEPARstart{H}{yperspectral Sensor} collects the information (reflectance) in several hundreds of contiguous bands with a very high spectral resolution which enables us to classify the objects based on their spectral signatures. However, these images are in relatively low spatial resolution due to the sensor limitations, SNR, and complexity constraints which significantly affect the performance for several real-world applications \cite{Shaohui17}. The traditional classifiers, for instance, KNN \cite{Ahmad19RS}, SVM \cite{Wang20}, Maximum Likelihood \cite{alcolea20}, Logistic Regression \cite{Ahmad19RS} and Extreme Learning Machine (ELM) \cite{Ahmad20Optik} are only works based on spectral information. These classifiers do not perform well due to spectral redundancy and high correlation among the spectral bands. Furthermore, these classifiers fail to preserve the important spatial variability of Hyperspectral data which also results in low performance.

The simplest way to improve the classification performance is to design a classifier that should incorporate both spectral and spatial information. Spatial information is considered as additional discriminative information associated with the size, shape, and structure of the object which if provided correctly brings more competitive results. Spatial spectral classifiers can generally be classified into two groups. First category explores the spatial and spectral information separately. The spatial information is extracted in advance using entropy \cite{Tuia14}, morphological operations \cite{Ghamisi15, Benediktsson05}, low rant representation \cite{Jia15} and attribute profiles \cite{Dalla11}. Later this information is combined with spectral information to perform pixel-level classification. 

The second category fuses the spatial-spectral information to get the joint features \cite{Zhong14}, for instance, 3D wavelet, scattering wavelet and Gabor filter \cite{Shen11, Qian13} are generated at different frequencies and scales to extract the joint spatial-spectral features for classification. Hyperspectral images are in 3D cubes thus the former category results in several 3D features i.e., spatial-spectra feature cubes comprising key information, thus preserving joint spatial-spectral correlations while extracting features can produce better results. However, the classical feature extraction methods are based on shallow learning and handcrafted features which largely depend on domain knowledge \cite{Swalpa19}. Therefore, the Deep models have been proposed to address the aforementioned issues i.e., automatically learn low to high-level features from raw HSI data which have attained incredible success for Hyperspectral Image Classification (HSIC). 

The last few years witnessed an intensive improvement in Convolutional Neural Network (CNN) for HSIC where the spatial features are tailored by a 2D CNN model \cite{Yuting20, Fang20, Huang20}. However, these spatial features are usually extracted separately which to some extent void the reason to jointly exploit the spatial-spectral information for HISC. Therefore, in this paper, a novel 3D CNN for HSIC method is proposed. This work first divides the HSI cube into small overlapping 3D patches. These patches are processed to generate 3D feature maps using 3D kernel function over multiple contiguous bands to preserve the joint spatial and spectral information for the feature learning process which exploits important discrimination information for HSIC. As a preprocessing, incremental Principle Component Analysis (iPCA) is deployed to reduce the redundancy among the bands to process the few important wavelengths out of the entire HSI cube. Later the 3D CNN classifier is trained in an end-to-end fashion which involves fewer parameters than other 2D/3D CNN models. The comparative study is also carried out with several state-of-the-art 2D/3D CNN based HSIC methods proposed in the literature. Experimental/comparative results revealed that the proposed method outperforms the compared ones.

The rest of the paper is structured as follows; section \ref{Sec.2} presents the proposed methodology. Section \ref{Sec.3} describes the experimental Datasets, Results and discussion. Finally Section \ref{Sec.4} concludes the paper with possible future research directions.

\section{Proposed Methodology}
\label{Sec.2}

\begin{figure}[!hbt]
    \centering
    \includegraphics[scale=0.07]{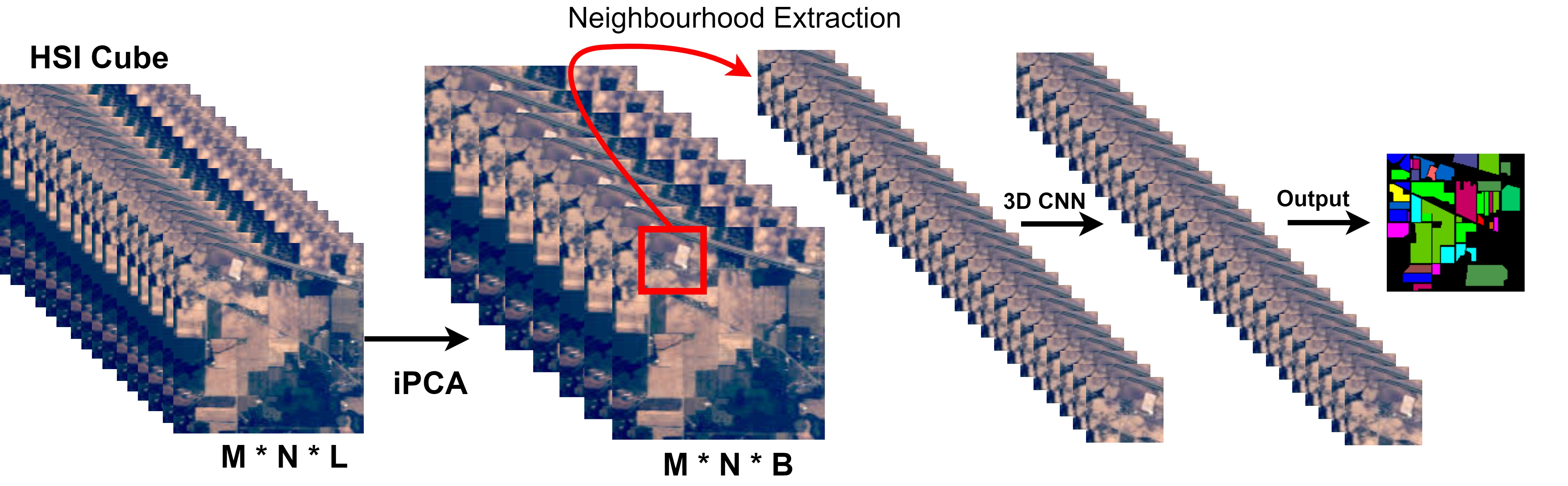}
    \caption{Proposed 3D CNN Model for HSIC. 3D CNN Model details, i.e., the number of 3D Convectional and fully connected layers, can be found in Table \ref{Tab.1}.}
    \label{Fig.1}
\end{figure}

Let us assume a Hyperspectral dataset can be expressed as $X=[x_1,x_2,x_3,...,x_L]^T \in R^{L \times(N \times M)}$ consisting of $N\times M$ samples associated with $C$ classes per band with total $L$ bands, in which each sample is represented as $(x_i,y_j)$, where $y_j$ is the class label of $x_i$ sample. In a nutshell $i^{th}$ sample belongs to $j^{th}$ class. Since the HSI pixels exhibit high inter-class similarity, high intra-class variability, overlapping, and nested regions which required intensive efforts to handle for any classification model \cite{Ahmad11, Ahmad112, Ahmad113, Ahmad122}. To overcome the aforesaid issues, incremental Principle Component Analysis (iPCA) is applied to the HSI cube to eliminate the redundant bands. iPCA reduces the number of images/bands ($L$ to $B$, where $B \ll L$) while maintaining the spatial dimensions as shown in Figure \ref{Fig.1}. 

\begin{table}[!hbt]
    \caption{Layer based Summary of our Proposed 3D CNN Model architecture shown in Figure \ref{Fig.2} with Window Size set as $11 \times 11$ for a sub-scene of Salinas Dataset.}
    \centering
    \begin{tabular}{c|c|c}  \hline
    \textbf{Layer} & \textbf{Output Shape} & \textbf{$\#$ of Parameters} \\ \hline 
    Input Layer & (11, 11, 20, 1) & 0 \\
    Conv3D\_1 (Conv3D) & (9, 9, 14, 8) & 512 \\  
    Conv3D\_2 (Conv3D) & (7, 7, 10, 16) & 5776  \\
    Conv3D\_3 (Conv3D) & (5, 5, 8, 32) & 13856 \\
    Conv3D\_4 (Conv3D) & (3, 3, 6, 64) & 55360 \\
    Flatten\_1 (Flatten) & (3456)& 0 \\
    Dense\_1 (Dense) & (256) & 884992 \\
    Dropout\_1 (Dropout) & (256) & 0 \\
    Dense\_2 (Dense) & (128) & 32896 \\
    Dropout\_2 (Dropout) & (128) & 0 \\ 
    Dense\_3 (Dense) & ($\#$ of Classes) & 774 \\ \hline
    \multicolumn{3}{c}{In total, \textbf{994,166} trainable parameters are required}  \\ \hline
    \end{tabular}
    \label{Tab.1}
\end{table}

\begin{figure}[!hbt]
    \centering
    \includegraphics[scale=0.10]{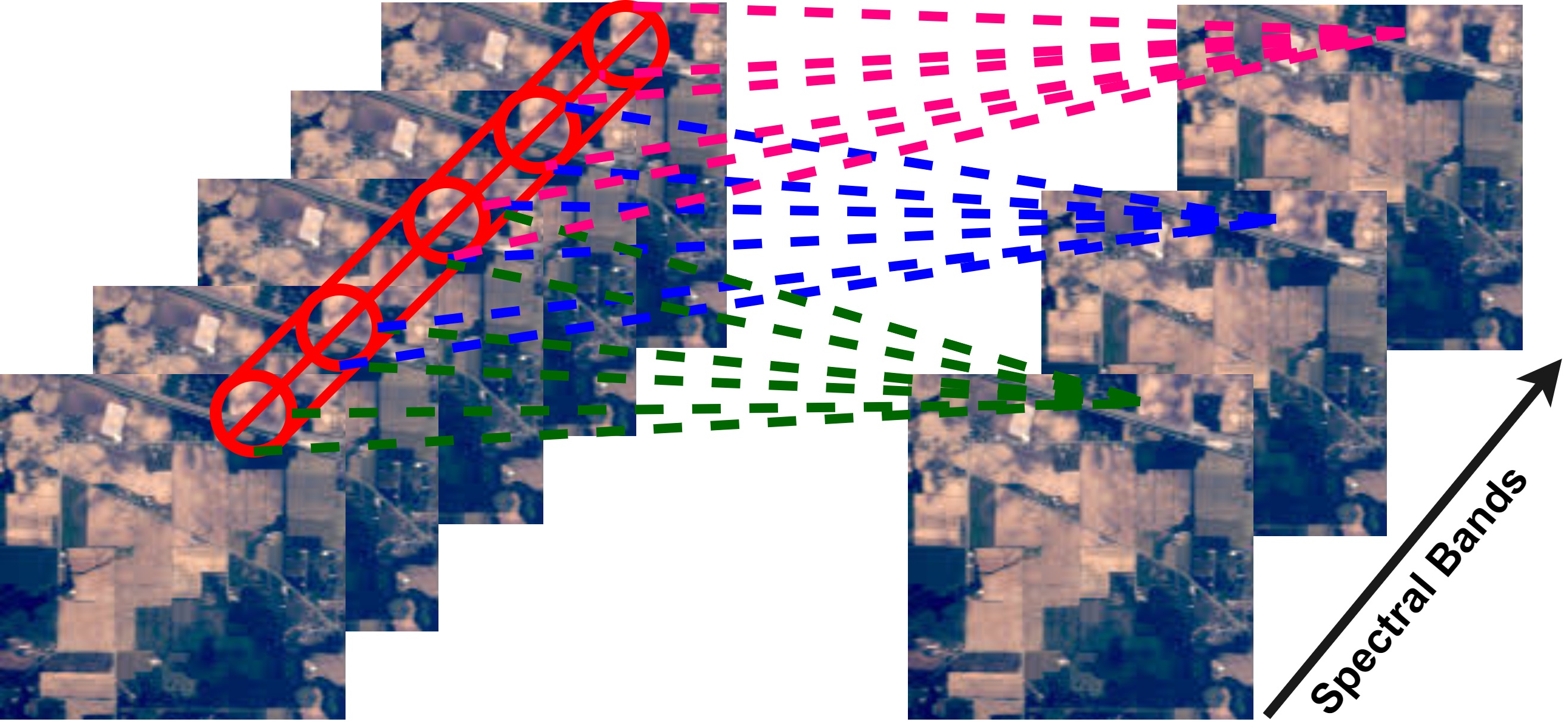}
    \caption{3D Convolution Operation}
    \label{Fig.2}
\end{figure}

In ordered to pass the HSI cube to the model, it must have to be divided into a small overlapping 3D spatial patches on which the ground labels are formed based on the central pixel as shown in Figure \ref{Fig.2}. The process creates neighboring patches $P \in R^{S \times S \times B}$ centered at the spatial location $(a, b)$ cover $(S \times S)$ spatial windows \cite{Swalpa19}. The the total of $n$ patches given by $(M – S + 1) \times (N – S + 1)$. Thus, these patches cover the width from $\frac{a - (S - 1)}{2}$ to $\frac{a + (S - 1)}{2}$ and height from $\frac{b - (S - 1)}{2}$ to $\frac{b + (S - 1)}{2}$.

The input patches are first convolved with 3D kernel function \cite{Ying17} which computes the sum of the dot product between kernel function and input patch \cite{Swalpa19}. Later these learned features are processed through an activation function that introduces the nonlinearity. In our proposed model, the activation values at spatial position $(x, y, z)$ in the $i^{th}$ layer and $j^{th}$ feature map is denoted as $v_{i,j}^{x,y,z}$, the the final model can be created as follows:

\begin{equation}
    v_{i,j}^{x,y,z} = \mathcal{F} \bigg(\sum_{\tau = 1}^{d_{t-1}} \sum_{\lambda = -\nu}^{\nu} \sum_{\rho = -\gamma}^{\gamma} \sum_{\phi = -\delta}^{\delta} w_{i, j, \tau}^{\nu, \rho, \lambda} \times v_{(i-1), \tau}^{(x+\nu), (y+\rho), (z+\lambda)} + b_{i,j} \bigg) 
\end{equation}
where $\mathcal{F}$ is an activation function, $d_{l - 1}$ be the number of 3D feature maps at $(l - 1)^{th}$ layer and $w_{i, j}$ be the depth of the kernel, $b_{i,j}$ is the bias, $2\delta +1$, $2\lambda +1$ and $2\nu + 1$ be the height, width and depth of the kernel.

In a nutshell, the proposed 3D CNN convolutional kernels are as follows: $3D\_conv\_layer1 = 8 \times 3 \times 3 \times 7 \times 1$ where $K_1^1 = 3, K_2^1 = 3$ and $K_3^1 = 7$. $3D\_conv\_layer2 = 16 \times 3\times 3\times 5 \times 8$ where $K_1^2 = 3, K_2^2 = 3$ and $K_3^2 = 5$.  $3D\_conv\_layer3 = 32 \times 3\times 3\times 3 \times 16$ where $K_1^3 = 3, K_2^3 = 3$ and $K_3^3 = 3$ and finally $3D\_conv\_layer4 = 64 \times 3 \times 3 \times 3 \times 16$ where $K_1^3 = 3, K_2^3 = 3$ and $K_3^3 = 3$. To increase the number of spatial-spectral feature maps, $4$ 3D convolutional layers are deployed before the flatten layer to make sure the model is able to discriminate the spatial information within different spectral bands without any loss. The further details regarding the proposed model can be found in Table \ref{Tab.1}. The total number of parameters (i.e., tune-able weights) of our proposed 3D CNN model is $994, 166$. The weights are initially randomized and optimized using Adam optimizer back-propagation with a soft-max loss function. The weights are updated using a mini-batch of size $256$ with $50$ epochs without batch normalization and augmentation.

\section{Experimental Datasets and Results}
\label{Sec.3}

The Salinas dataset (SD) was acquired over Salinas Valley California using AVIRIS sensor. SD is of size $512 \times 217 \times 224$ with a $3.7$ meter spatial resolution with $512 \times 217$ is spatial and $224$ spectral dimensions. SD consists of vineyard fields, vegetables and bare soils. SD consist of $16$ classes. A few water absorption bands $108-112, 154-167$ and $224$ are removed before analysis.

Indian Pines  Dataset (IPD) is obtained over northwestern Indiana’s test site by Airborne Visible / Infrared Imaging Spectrometer (AVIRIS) sensor. IPD is of size $145 \times 145 \times 224$ in the wavelength range $0.4-2.5 \times 10^{-6}$ meters where $145 \times 145$ is the spatial and $224$ spectral dimensions. IPD consists of $1/3$ forest and $2/3$ agriculture area and other naturally evergreen vegetation. Some corps in the early stages of their growth is also present with approximately less than $5\%$ of total coverage. Low-density housing, building and small roads, Two dual-lane highway and a railway line are also a part of IPD. The IPD ground truth comprised of $16$ classes which are not mutually exclusive. The water absorption bands have been removed before the experiments thus the remaining $200$ bands are used in this experiment. 

\begin{figure}[H]
	\begin{subfigure}{0.09\textwidth}
		\includegraphics[scale=0.3]{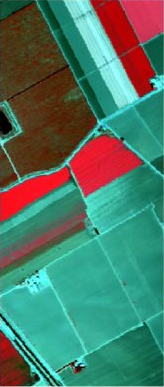}
		\centering
		\caption{SA} 
		\label{Fig.3A}
	\end{subfigure}
	\begin{subfigure}{0.10\textwidth}
		\includegraphics[scale=0.19]{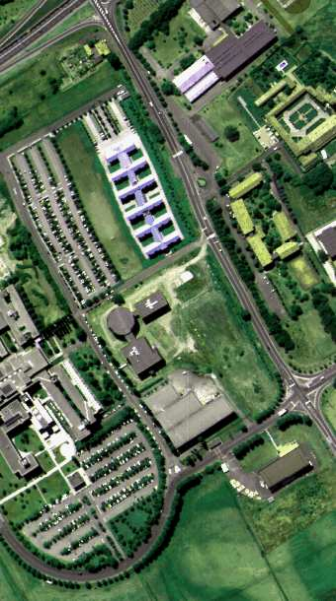}
		\centering
		\caption{PU} 
		\label{Fig.3B}
	\end{subfigure}
	\begin{subfigure}{0.13\textwidth}
		\includegraphics[scale=0.3]{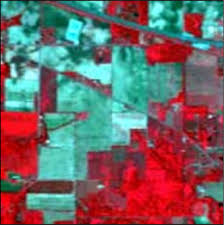}
		\centering
		\caption{IP} 
		\label{Fig.3C}
	\end{subfigure}
	\begin{subfigure}{0.13\textwidth}
		\includegraphics[scale=0.48]{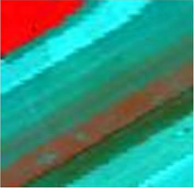}
		\centering
		\caption{SL-A} 
		\label{Fig.3D}
	\end{subfigure}
\caption{Ground images of experimental datasets used in this work.}
\label{Fig.3}
\end{figure}

Pavia University Dataset (PUD) gathered over Pavia in northern Italy using a Reflective Optics System Imaging Spectrometer (ROSIS) optical sensor. PUD consists of $610 \times 610$ spatial and $103$ spectral bands with a spatial resolution of $1.3$ meters. PUD ground truth classes are $9$. Further details about the experimental datasets can be found at \cite{Grupointel}. The ground images of all the experimental datasets are shown in Figure \ref{Fig.3}.

All the experiments were performed on an online platform known as Google Colab \cite{carneiro2018}. Google Colab is an online platform that requires a good speed of internet to run any environment. Google Colab provides an option to execute the codes on python $3$ notebook with Graphical Processing Unit (GPU), $25$ GB of Random Access Memory (RAM) and $358.27$ GB of could storage for data computation. In all the experiments, the initial Test/Train set is divided into a $30/70\%$ ratio on which Training samples ($70\%$ of the entire population) are further divided into $50/50\%$ for the Training and Validation set. 

To make the fair comparisons, the learning rate for all the experiments is set to $0.001$, $relu$ as an activation function is used for all layers except last on which $softmax$ is used, the patch sizes are set as $11 \times 11 \times 20$, $13 \times 13 \times 20$, $15 \times 15 \times 20$, $17 \times 17 \times 20$, $19 \times 19 \times 20$, $21 \times 21 \times 20$ and $25 \times 25 \times 20$, respectively with $20$ most informative bands selected by iPCA method. For evaluation purposes, Average Accuracy (AA), Overall Accuracy (OA) and Kappa ($\kappa$) coefficient have been computed form the confusion matrices. AA represents the average class-wise classification performance, OA is computed as the number of correctly classified examples out of the total test examples and finally, $\kappa$ is known as a statistical metric that considered the mutual information regarding a strong agreement among classification and ground-truth maps. Along with OA, AA and $\kappa$ metrics, several statistical tests are also being considered such as F1-Score, Precision and Recall. 

\begin{figure}[!hbt]
	\begin{subfigure}{0.24\textwidth}
		\includegraphics[scale=0.25]{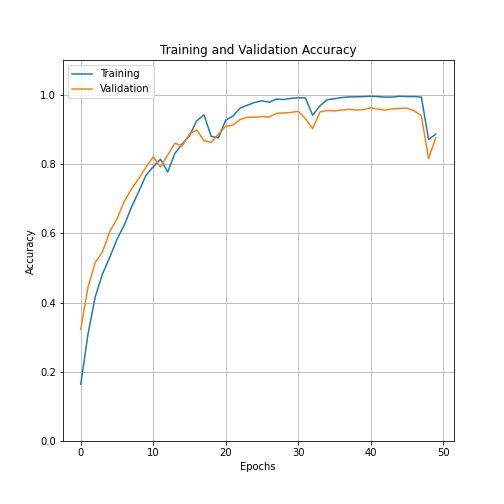}
		\centering
		\caption{Accuracy} 
		\label{Fig.4A}
	\end{subfigure}
	\begin{subfigure}{0.24\textwidth}
		\includegraphics[scale=0.25]{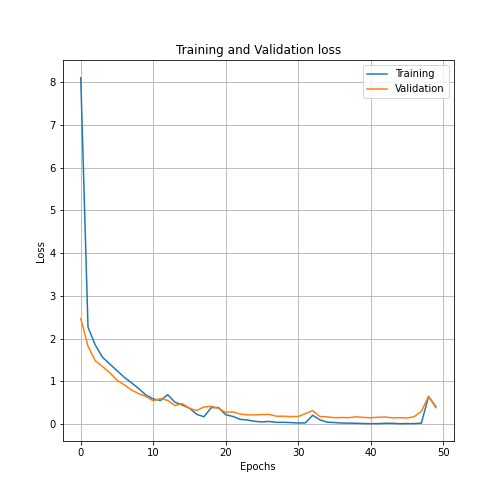}
		\centering
		\caption{Loss} 
		\label{Fig.4B}
	\end{subfigure}
\caption{Accuracy and Loss for Training and Validation sets on Indian Pines Dataset with $11 \times 11$ window patch corresponds to the $50$ number of Epochs.}
\label{Fig.4}
\end{figure}

The convergence loss and accuracy of our proposed 3D CNN model for a $50$ number of epochs are shown in Figure \ref{Fig.4}. From these figures, one can conclude that the proposed model is converged almost around $32$ echos.
Whereas, the computational time of our proposed model is shown in Table \ref{Tab.2} which also reveals a fast convergence and computational efficiency of our proposed model. The computational time highly depends on the speed of the internet and available RAM.
\begin{table}[!hbt]
    \centering
    \caption{Computational time in minutes for all the experimental datasets with several window sizes.}
    \resizebox{\columnwidth}{!}{
    \begin{tabular}{c|c|c|c|c|c|c|c} \hline
        \multirow{2}{*}{\textbf{Dataset}} & \multicolumn{7}{c}{\textbf{Window Size}} \\
        & $11 \times 11$ & $13 \times 13$ & $15 \times 15$ & $17 \times 17$ & $19 \times 19$ & $21 \times 21$ & $25 \times 25$ \\ \hline 
        SL-A & 0.22 & 0.23 & 0.56 & 0.28 & 0.98 & 0.37 & 0.45 \\
        SL & 1.34 & 1.41 & 1.60 & 2.00 & 3.17 & 2.63 & 3.52 \\
        IP & 0.33 & 0.33 & 0.61 & 0.78 & 0.62 & 0.58 & 0.76 \\ 
        PU & 2.16 & 5.26 & 1.35 & 2.00 & 2.46 & 2.14 & 2.83 \\ \hline 
    \end{tabular} 
    }
    \label{Tab.2}
\end{table}

The accuracy analysis i.e., OA, AA, and $\kappa$ based on the impact of spatial dimensions \footnote{The Confusion matrices (with per class accuracy) for each window size and every dataset is provided in the supplementary material.} processed by the proposed model is presented in Table \ref{Tab.3}. While looking into the Table \ref{Tab.3}, one can conclude that the window size of $11 \times 11$ is enough for Pavia University, Salinas and Salinas-A dataset whereas the window size of $13 \times 13$ and $25 \times 25$ both works almost the same. 

Furthermore, the classification maps (geographical locations for each class) according to the different number of window sizes (spatial dimensions) are shown in Figures \ref{Fig.5}-\ref{Fig.8}. In regards to comparison, the proposed model is compared with several state-of-the-art methods published in the recent few years. From experimental results listed in Table \ref{Tab.4} one can conclude that the proposed model has competitive results and to some extent better in regards to the other methods. The comparative methods includes Multi-scale-3D-CNN \cite{He17}, 3D/2D-CNN \cite{Ying17, Luo18, Hamida18, Bing17}. 
\begin{table}[!hbt]
    \centering
    \caption{Impact of window size on our proposed model}
    \resizebox{\columnwidth}{!}{
    \begin{tabular}{c|c|c|c|c|c|c|c|c|c|c|c|c} \hline 
    \multirow{2}{*}{\textbf{Window}} & \multicolumn{3}{c|}{\textbf{PU}} & \multicolumn{3}{c|}{\textbf{IP}} & \multicolumn{3}{c|}{\textbf{SA}} & \multicolumn{3}{c}{\textbf{SL-A}} \\
    & \textbf{OA} & \textbf{AA} & \textbf{$\kappa$} & \textbf{OA} & \textbf{AA} & \textbf{$\kappa$} & \textbf{OA} & \textbf{AA} & \textbf{$\kappa$} & \textbf{OA} & \textbf{AA} & \textbf{$\kappa$} \\ \hline 
    $11 \times 11$ & 99.94 & 99.89 & 99.92 & 88.65 & 83.52 & 87.11 & 99.80 & 99.91 &  99.78 & 100 & 100 & 100 \\
    $13 \times 13$ & 99.81 & 99.65 & 99.75 & 95.38 & 94.14 & 94.72 & 99.93 & 99.94 & 99.93 & 100 & 100 & 100 \\
    $15 \times 15$ & 99.85 & 99.62 & 99.80 & 93.69 & 93.09 & 92.79 & 99.99 & 99.99 & 99.99 & 100 & 100 & 100 \\
    $17 \times 17$ & 99.05 & 98.49 & 98.75 & 91.80 & 91.74 & 90.62 & 99.95 & 99.97 & 99.95 & 99.93 & 99.93 & 99.92 \\
    $19 \times 19$ & 99.93 & 99.78 & 99.91 & 93.13 & 93.42 & 92.15 & 98.04 & 94.02 & 97.81 & 100 & 100 & 100 \\
    $21 \times 21$ & 99.78 & 99.43 & 99.72 & 94.34 & 91.31 & 93.52 & 99.99 & 99.99 & 99.99 & 100 & 100 & 100 \\
    $25 \times 25$ & 98.79 & 97.67 & 98.39 & 97.75 & 96.17 & 97.44 & 99.96 & 99.93 & 99.95 & 100 & 100 & 100 \\ \hline 
    \end{tabular}
    }
    \label{Tab.3}
\end{table}
From experiments listed in Table \ref{Tab.4} shows the proposed method improves the results significantly then the state-of-the-art methods with even less number of training samples. 
\begin{table}[!hbt]
    \centering
    \caption{Comparative evaluations with State-of-the-art methods while considering $11 \times 11$ Spatial dimensions and $10\%$ of training samples.}
    \resizebox{\columnwidth}{!}{
    \begin{tabular}{c|c|c|c|c|c|c|c|c|c|c|c|c} \hline 
    \multirow{2}{*}{dataset} & \multicolumn{3}{c|}{Multi-scale-3D-CNN} & \multicolumn{3}{c|}{3D-CNN} &  \multicolumn{3}{c|}{2D-CNN} & \multicolumn{3}{c}{\textbf{Proposed}} \\
    & OA  & AA  & Kappa  & OA  & AA  & Kappa & OA  & AA  & Kappa & OA  & AA  & Kappa  \\ \hline 
    PU & 95.95 & 97.52 & 93.40 &  96.34 & 97.03 & 94.90 & 96.63 & 94.84 & 95.53 & \textbf{98.40} & \textbf{97.89} & \textbf{97.89}\\ 
    IP & 81.39 & 75.22 & 81.20 & 82.62 & 76.51 & 79.25 & 80.27 & 68.32 & 75.26 & \textbf{97.75} & \textbf{94.54} & \textbf{97.44} \\
    SA & 94.20 & 96.66 & 93.61 & 85.00 & 89.63 & 83.20 & 96.34 & 94.36 & 95.93 &  \textbf{98.06} & \textbf{98.80} & \textbf{97.85}\\ \hline 
    \end{tabular}
    }
    \label{Tab.4}
\end{table}

\begin{figure*}[!hbt]
	\begin{subfigure}{0.12\textwidth}
		\includegraphics[scale=0.40]{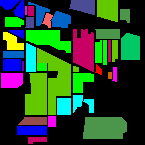}
		\centering
		\caption{GT} 
		\label{Fig.7A}
	\end{subfigure}
	\begin{subfigure}{0.12\textwidth}
		\includegraphics[scale=0.40]{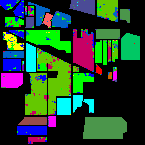}
		\centering
		\caption{$11 \times 11$} 
		\label{Fig.7B}
	\end{subfigure}
	\begin{subfigure}{0.12\textwidth}
		\includegraphics[scale=0.40]{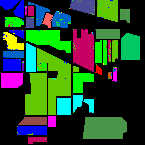}
		\centering
		\caption{$13 \times 13$} 
		\label{Fig.7C}
	\end{subfigure}
	\begin{subfigure}{0.12\textwidth}
		\includegraphics[scale=0.40]{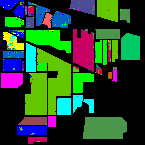}
		\centering
		\caption{$15 \times 15$} 
		\label{Fig.7D}
	\end{subfigure}
	\begin{subfigure}{0.12\textwidth}
		\includegraphics[scale=0.40]{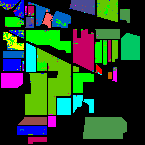}
		\centering
		\caption{$17 \times 17$} 
		\label{Fig.7E}
	\end{subfigure}
	\begin{subfigure}{0.12\textwidth}
		\includegraphics[scale=0.40]{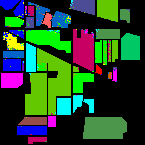}
		\centering
		\caption{$19 \times 19$} 
		\label{Fig.7F}
	\end{subfigure}
	\begin{subfigure}{0.12\textwidth}
		\includegraphics[scale=0.40]{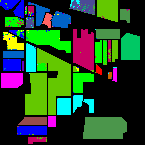}
		\centering
		\caption{$21 \times 21$} 
		\label{Fig.7G}
	\end{subfigure}	
	\begin{subfigure}{0.12\textwidth}
		\includegraphics[scale=0.40]{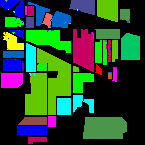}
		\centering
		\caption{$25 \times 25$} 
		\label{Fig.7H}
	\end{subfigure}
\caption{\textbf{Indian Pines Dataset} Ground Truths for each spatial dimensions processed through our proposed model.}
\label{Fig.7}
\end{figure*}

\begin{figure*}[!hbt]
	\begin{subfigure}{0.12\textwidth}
		\includegraphics[scale=0.25]{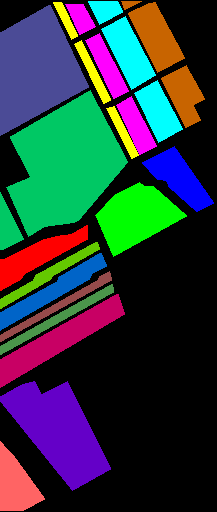}
		\centering
		\caption{GT} 
		\label{Fig.5A}
	\end{subfigure}
	\begin{subfigure}{0.12\textwidth}
		\includegraphics[scale=0.25]{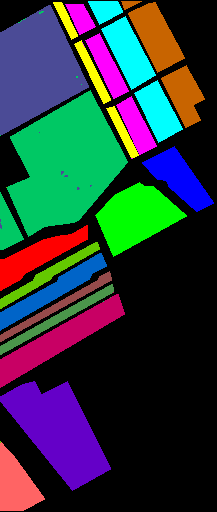}
		\centering
		\caption{$11 \times 11$} 
		\label{Fig.5B}
	\end{subfigure}
	\begin{subfigure}{0.12\textwidth}
		\includegraphics[scale=0.25]{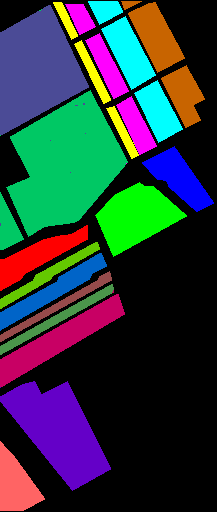}
		\centering
		\caption{$13 \times 13$} 
		\label{Fig.5C}
	\end{subfigure}
	\begin{subfigure}{0.12\textwidth}
		\includegraphics[scale=0.25]{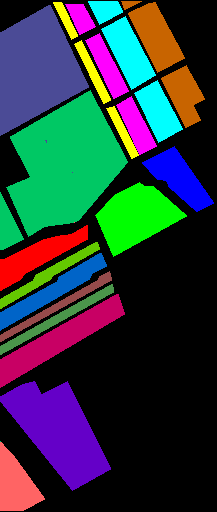}
		\centering
		\caption{$15 \times 15$} 
		\label{Fig.5D}
	\end{subfigure}
	\begin{subfigure}{0.12\textwidth}
		\includegraphics[scale=0.25]{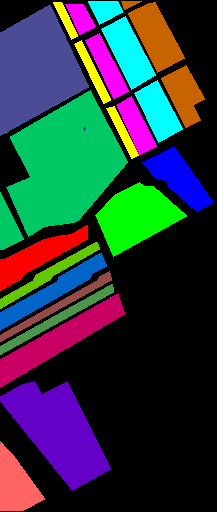}
		\centering
		\caption{$17 \times 17$} 
		\label{Fig.5E}
	\end{subfigure}
	\begin{subfigure}{0.12\textwidth}
		\includegraphics[scale=0.25]{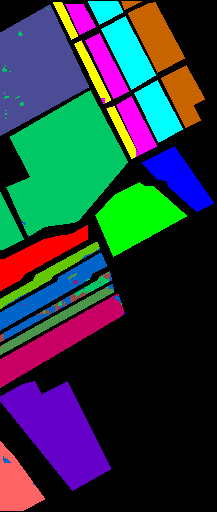}
		\centering
		\caption{$19 \times 19$} 
		\label{Fig.5F}
	\end{subfigure}
	\begin{subfigure}{0.12\textwidth}
		\includegraphics[scale=0.25]{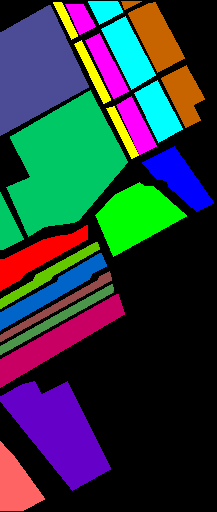}
		\centering
		\caption{$21 \times 21$} 
		\label{Fig.5G}
	\end{subfigure}	
	\begin{subfigure}{0.12\textwidth}
		\includegraphics[scale=0.25]{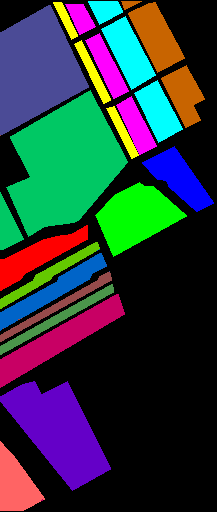}
		\centering
		\caption{$25 \times 25$} 
		\label{Fig.5H}
	\end{subfigure}
\caption{\textbf{Salinas Dataset} Ground Truths for each spatial dimensions processed through our proposed model.}
\label{Fig.5}
\end{figure*}

\begin{figure*}[!hbt]
	\begin{subfigure}{0.12\textwidth}
		\includegraphics[scale=0.18]{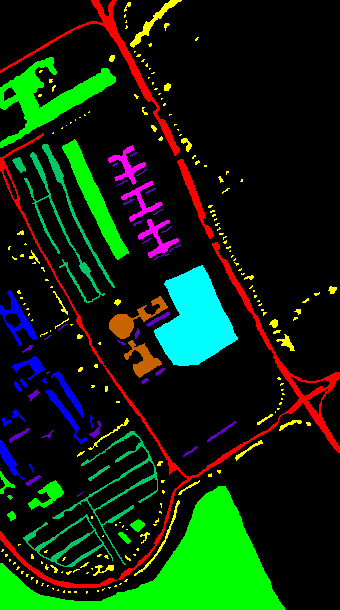}
		\centering
		\caption{GT} 
		\label{Fig.6A}
	\end{subfigure}
	\begin{subfigure}{0.12\textwidth}
		\includegraphics[scale=0.18]{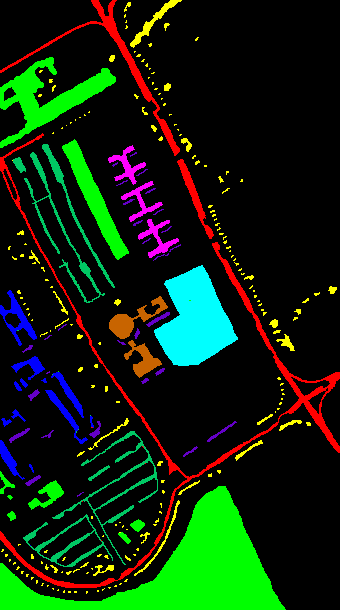}
		\centering
		\caption{$11 \times 11$} 
		\label{Fig.6B}
	\end{subfigure}
	\begin{subfigure}{0.12\textwidth}
		\includegraphics[scale=0.18]{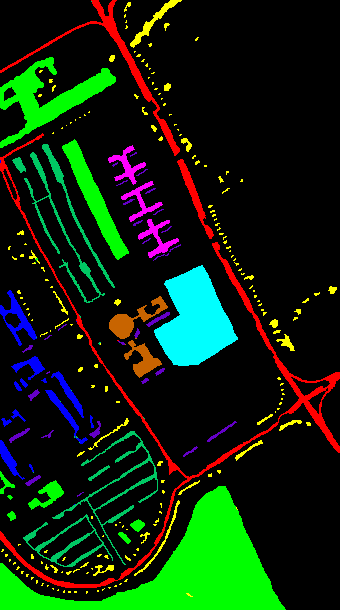}
		\centering
		\caption{$13 \times 13$} 
		\label{Fig.6C}
	\end{subfigure}
	\begin{subfigure}{0.12\textwidth}
		\includegraphics[scale=0.18]{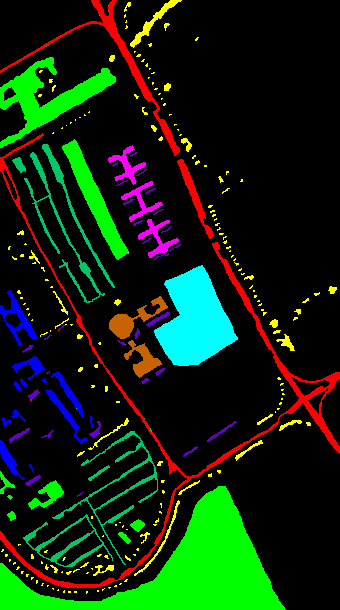}
		\centering
		\caption{$15 \times 15$} 
		\label{Fig.6D}
	\end{subfigure}
	\begin{subfigure}{0.12\textwidth}
		\includegraphics[scale=0.18]{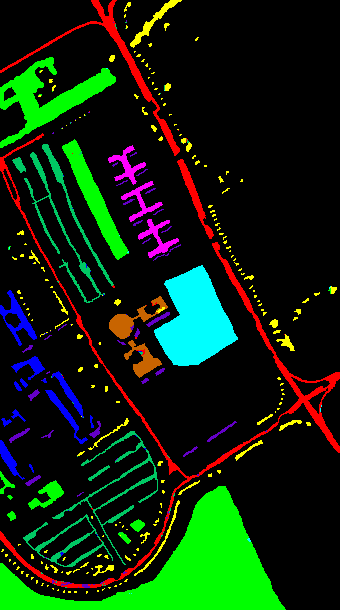}
		\centering
		\caption{$17 \times 17$} 
		\label{Fig.6E}
	\end{subfigure}
	\begin{subfigure}{0.12\textwidth}
		\includegraphics[scale=0.18]{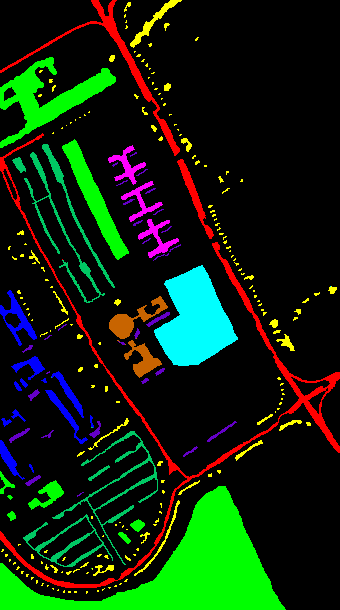}
		\centering
		\caption{$19 \times 19$} 
		\label{Fig.6F}
	\end{subfigure}
	\begin{subfigure}{0.12\textwidth}
		\includegraphics[scale=0.18]{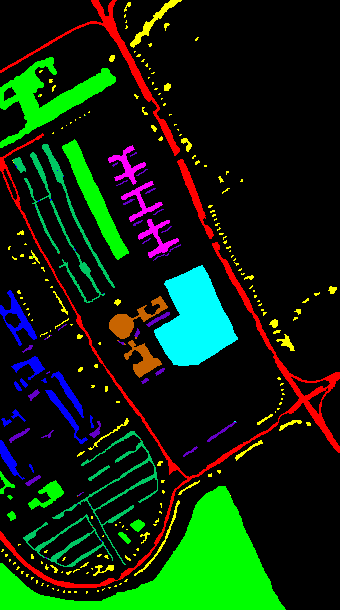}
		\centering
		\caption{$21 \times 21$} 
		\label{Fig.6G}
	\end{subfigure}	
	\begin{subfigure}{0.12\textwidth}
		\includegraphics[scale=0.18]{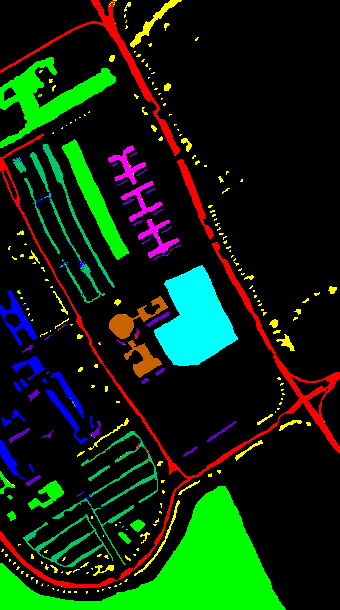}
		\centering
		\caption{$25 \times 25$} 
		\label{Fig.6H}
	\end{subfigure}
\caption{\textbf{Pavia University Dataset} Ground Truths for each spatial dimensions processed through our proposed model.}
\label{Fig.6}
\end{figure*}

\begin{figure*}[!hbt]
	\begin{subfigure}{0.12\textwidth}
		\includegraphics[scale=0.65]{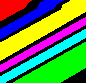}
		\centering
		\caption{GT} 
		\label{Fig.8A}
	\end{subfigure}
	\begin{subfigure}{0.12\textwidth}
		\includegraphics[scale=0.65]{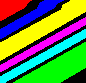}
		\centering
		\caption{$11 \times 11$} 
		\label{Fig.8B}
	\end{subfigure}
	\begin{subfigure}{0.12\textwidth}
		\includegraphics[scale=0.65]{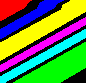}
		\centering
		\caption{$13 \times 13$} 
		\label{Fig.8C}
	\end{subfigure}
	\begin{subfigure}{0.12\textwidth}
		\includegraphics[scale=0.65]{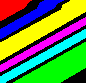}
		\centering
		\caption{$15 \times 15$} 
		\label{Fig.8D}
	\end{subfigure}
	\begin{subfigure}{0.12\textwidth}
		\includegraphics[scale=0.65]{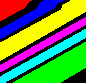}
		\centering
		\caption{$17 \times 17$} 
		\label{Fig.8E}
	\end{subfigure}
	\begin{subfigure}{0.12\textwidth}
		\includegraphics[scale=0.65]{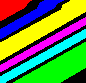}
		\centering
		\caption{$19 \times 19$} 
		\label{Fig.8F}
	\end{subfigure}
	\begin{subfigure}{0.12\textwidth}
		\includegraphics[scale=0.65]{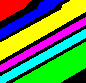}
		\centering
		\caption{$21 \times 21$} 
		\label{Fig.8G}
	\end{subfigure}	
	\begin{subfigure}{0.12\textwidth}
		\includegraphics[scale=0.65]{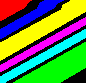}
		\centering
		\caption{$25 \times 25$} 
		\label{Fig.8H}
	\end{subfigure}
\caption{\textbf{Salinas-A Dataset} Ground Truths for each spatial dimensions processed through our proposed model.}
\label{Fig.8}
\end{figure*}

\section{Conclusion}
\label{Sec.4}
Hyperspectral Image Classification (HSIC) is a challenging task due to high inter-class similarity and high intra-class variability. Therefore, this paper proposed a lightweight fast 3D CNN model which not only overcome the abovesaid challenges but also provide state of the art experimental results in a computationally efficient fashion on three benchmark Hyperspectral datasets. The inter-class similarity and high intra-class variability issues are being resolved using a spatial-spectral information using 3D convolutions. The experimental results reveal that the proposed method outperformed the state of the art methods, furthermore, the proposed model is less complex than the conventional 3D CNN models. 
\section*{Running Code}
The Running Demo can be found at \href{https://github.com/mahmad00/A-Fast-3D-CNN-for-HSIC}{Github}.
{\footnotesize
\bibliographystyle{IEEEtran}
\bibliography{Sample}
}
\end{document}